%% file: ms.tex
\documentclass[conference]{IEEEtran}
\pdfoutput=1
\usepackage{cite}
\usepackage{amsmath,amssymb,amsfonts}
\usepackage{algorithmic}
\usepackage{graphicx}
\usepackage{xcolor}

\usepackage{booktabs} 
\usepackage{subfigure}
\usepackage{amsmath}
\usepackage[normalem]{ulem}
\usepackage{enumitem}
\usepackage{multirow}
\usepackage[nomargin,inline,marginclue,draft]{fixme}
\usepackage{balance}
\usepackage{changepage}

\newlength\savedwidth

\newcommand{\para}[1]{{\vspace{4pt} \bf \noindent #1 \hspace{10pt}}}

\newcommand{\davy}[1]{\textcolor[rgb]{0.,0.,0.}{#1}}

\def\BibTeX{{\rm B\kern-.05em{\sc i\kern-.025em b}\kern-.08em
    T\kern-.1667em\lower.7ex\hbox{E}\kern-.125emX}}
\begin{document}

\title{Price-aware Recommendation with Graph Convolutional Networks}

\author{\IEEEauthorblockN{Yu~Zheng\IEEEauthorrefmark{1},
Chen~Gao\IEEEauthorrefmark{1}, 
Xiangnan~He\IEEEauthorrefmark{2},
Yong~Li\IEEEauthorrefmark{1},
Depeng~Jin\IEEEauthorrefmark{1}
 }
 \IEEEauthorblockA{\IEEEauthorrefmark{1}Beijing National Research Center for Information Science and Technology\\
 Department of Electronic Engineering, Tsinghua University, Beijing 100084, China\\}
 \IEEEauthorblockA{\IEEEauthorrefmark{2}School of Information Science and Technology, University of Science and Technology of China\\
}
liyong07@tsinghua.edu.cn
} %

\maketitle

\begin{abstract}
In recent years, much research effort on recommendation has been devoted to mining user behaviors, i.e., collaborative filtering, along with the general information which describes users or items, e.g., textual attributes, categorical demographics, product images, and so on. Price, an important factor in marketing --- which determines whether a user will make the final purchase decision on an item --- surprisingly, has received relatively little scrutiny. 

In this work, we aim at developing an effective method to predict user purchase intention with the focus on the price factor in recommender systems. The main difficulties are two-fold: 1) the preference and sensitivity of a user on item price are unknown, which are only implicitly reflected in the items that the user has purchased, and 2) how the item price affects a user's intention depends largely on the product category, that is, the perception and affordability of a user on item price could vary significantly across categories. 
Towards the first difficulty, we propose to model the transitive relationship between user-to-item and item-to-price, taking the inspiration from the recently developed Graph Convolution Networks (GCN). The key idea is to propagate the influence of price on users with items as the bridge, so as to make the learned user representations be price-aware. For the second difficulty, we further integrate item categories into the propagation progress and model the possible pairwise interactions for predicting user-item interactions. We conduct extensive experiments on two real-world datasets, demonstrating the effectiveness of our GCN-based method in learning the price-aware preference of users. Further analysis reveals that modeling the price awareness is particularly useful for predicting user preference on items of unexplored categories. 
\end{abstract}

\begin{IEEEkeywords}
Price-aware, recommendation, collaborative filtering, price, user preference
\end{IEEEkeywords}

\input{1.introduction}
\input{2.preliminary}
\input{3.probdef}

\input{4.method}

\input{7.implementation.tex}
\input{5.experiments.1}
\input{6.relatedworks}

\section{Conclusion and Future Work}\label{Sec:Conclusion}
In this work, we highlight the significance of incorporating price into recommendation. To capture the two difficulties of incorporating price which are unstated price awareness and category-dependent influence, we propose a GCN-based method named PUP and adopt a two-branch structure which is specifically designed to disentangle the global and local effect of the price awareness.
We conduct extensive experiments on real-world datasets, demonstrating that our proposed PUP could improve the recommendation performance over existing methods. Further insights are provided on alleviating the cold start issue via capturing price awareness.

\davy{Although specifically designed for modeling price sensitivity, our proposed model serves great generality with respect to feature engineering where other features can be easily integrated into our proposed method. For example, user profiles can be added as separate nodes linked to user nodes, while item features other than price and category can be integrated similarly.} Empirical studies are needed to investigate the performance of PUP when further incorporating large scale features. As increasing research focusing on the price factor from the perspective of service providers, how to utilize PUP to maximize the revenue is an interesting and important research question which extends price-aware recommendation to value-aware recommendation. Furthermore, modeling the dynamic of price is also a promising future direction.

\section*{Acknowledgement}
This work was supported in part by The National Key Research and Development Program of China under grant SQ2018YFB180012, the National Nature Science Foundation of China under 61971267, 61972223, 61861136003, and 61621091, Beijing Natural Science Foundation under L182038, Beijing National Research Center for Information Science and Technology under 20031887521, and research fund of Tsinghua University - Tencent Joint Laboratory for Internet Innovation Technology.

\bibliographystyle{IEEEtran}
\bibliography{bibliography}
\balance

\end{document}

%% file: 1.introduction.tex
\section{Introduction}\label{Sec:Intro}
Recommendation is attracting increasing attention in both industry and academia, owing to the prevalence and success of recommender systems in many applications~\cite{eksombatchai2018pixie, zhou2018deep, covington2016deep}. 
From the perspective of product providers, the aim of building a recommender system is to increase the traffic and revenue, by recommending the items that a user will be likely to consume. As such, the key data source to leverage is the past consumption histories of users, since they provide direct evidence on a user's interest. To this end, much research effort has been devoted to collaborative filtering (CF)~\cite{kabbur2013fism, BPR, ncf}, which casts the task as completing the user-item consumption matrix, and incorporating side information into CF, such as textual attributes~\cite{wan2018one,zheng2017joint, wang2017item}, categorical demographics~\cite{chen2017attentive}, and product images~\cite{cheng2019mmalfm}. 
To utilize such diverse data in an unified model, a general class of feature-based recommendation models have been proposed, such as the pioneer work of factorization machines (FM)~\cite{FM} and several recent developments that augment FM with neural networks~\cite{DeepFM,xDeepFM,NFM}. 

With respect to E-commerce products and restaurants recommendation, where the item comes at an economic cost, not only the inherent interest of the user, but also the item price, plays a critical role in determining whether the user will make the final purchase decision. 
It has long been acknowledged that price is a significant factor in affecting user behaviors and product sales in marketing research~\cite{chen1998effects,krishnamurthi1992asymmetric}. Nevertheless, and surprisingly, it has received relatively little scrutiny in recommendation. 

Different from other item attributes like manufacturer and tags that influence a user's interest, the price of an item instead affects whether the user is willing to pay (WTP) for it. In other words, price and other attributes play orthogonal roles in the user decision making process --- in most cases, only when both the item is of interest and its price is acceptable, will the user purchase it. In general, there are two difficulties in effectively integrating item price into recommender systems:
\begin{itemize}[leftmargin=*]
    \item \textbf{Unstated price awareness.} A user seldom states her preference and sensitivity on item price explicitly. 
    As such, to build data-driven approaches, we have to infer a user's personalized awareness on item price from her purchase history. 
    More challengingly, we need to consider the CF effect reflected in the histories of similar users to enhance the inference accuracy. 
    \item \textbf{Category-dependent influence.} A user would have rather different perception and affordability on items of different categories. For example, a sport lover would have high tolerance on the price of a sport equipment, but not on alcoholic drinks. As such, it is important to take the item category information into consideration to accurately infer users' price preference.
\end{itemize}

\davy{As a special case of item side information, price could be integrated to generic recommender models like FM as a normalized numerical feature or discretized categorical feature.  
However, such solutions ignore the unique role of price in affecting user decision --- they are not specifically designed to tackle the above-mentioned two challenges, thus whether the price sensitivity is properly captured remains unclear.} In this work, we aim to address the two difficulties in price-aware recommendation system. We propose a new solution named \textit{\textbf{P}rice-aware \textbf{U}ser \textbf{P}reference-modeling} (PUP), which employs the recently emerged Graph Convolution Networks (GCN)~\cite{NGCF19} to learn the price-aware and category-dependent user representations. 

To be specific, we discretize the price variable and build a heterogeneous graph consisting of four types of nodes --- users, items, prices and categories --- where users connect to items, and items connect to prices and categories. 
We then propagate embeddings from prices to users with items as the bridge, so as to encode the indirect influence of prices on users. 
This makes the user embedding be related to the price embedding, such that high affinity is conceptually assigned to a user and her frequently purchased prices. 
To capture the CF effect of collective behaviors, we further propagate the user embeddings back to items and prices. 
Towards the second challenge of category relevance, we also integrate categories into the propagation process, and employ a pairwise interaction-based decoder to capture the interactions among users, items, prices and categories. 
Lastly, the overall model is trained in an end-to-end fashion and is optimized to estimate the consumption behaviors.  
Through these designs, our PUP method effectively incorporates the important yet complicated price factor into recommender systems.

To summarize, the main contributions of this work are as follows.
\begin{itemize}[leftmargin=*]
\item We highlight the significance of the price factor in recommending items with economic cost, and propose a graph-based solution to unify the influence of item price and category to learn user preference. 
\item We experiment on real-world datasets to evaluate our method. Further analysis justifies the utility of modeling price in cold-start scenarios and  recommending the items of unexplored categories for a user. 
\end{itemize}

The remainder of the paper is as follows. First we conduct some preliminary studies on a real-world dataset to analyze users' category-dependent price awareness in Section~\ref{Sec:Pre}. Then we formalize the problem in Section~\ref{Sec:ProbDef} and present our proposed method in Section~\ref{Sec:Method}. After that we introduce the implementation of our proposed method in Section~\ref{Sec:Implementation}. We conduct experiments in Section~\ref{Sec:Exp} and we review related work in Section~\ref{Sec:RelatedWork}. Last, we conclude the paper in Section~\ref{Sec:Conclusion}.

%% file: 2.preliminary.tex
\section{Motivation and Problem Formulation}
\subsection{Preliminary Study}\label{Sec:Pre}
\begin{figure}[t]
    \begin{minipage}[t]{1.0\linewidth}
        \includegraphics[width=\linewidth,bb=0 0 1440 864]{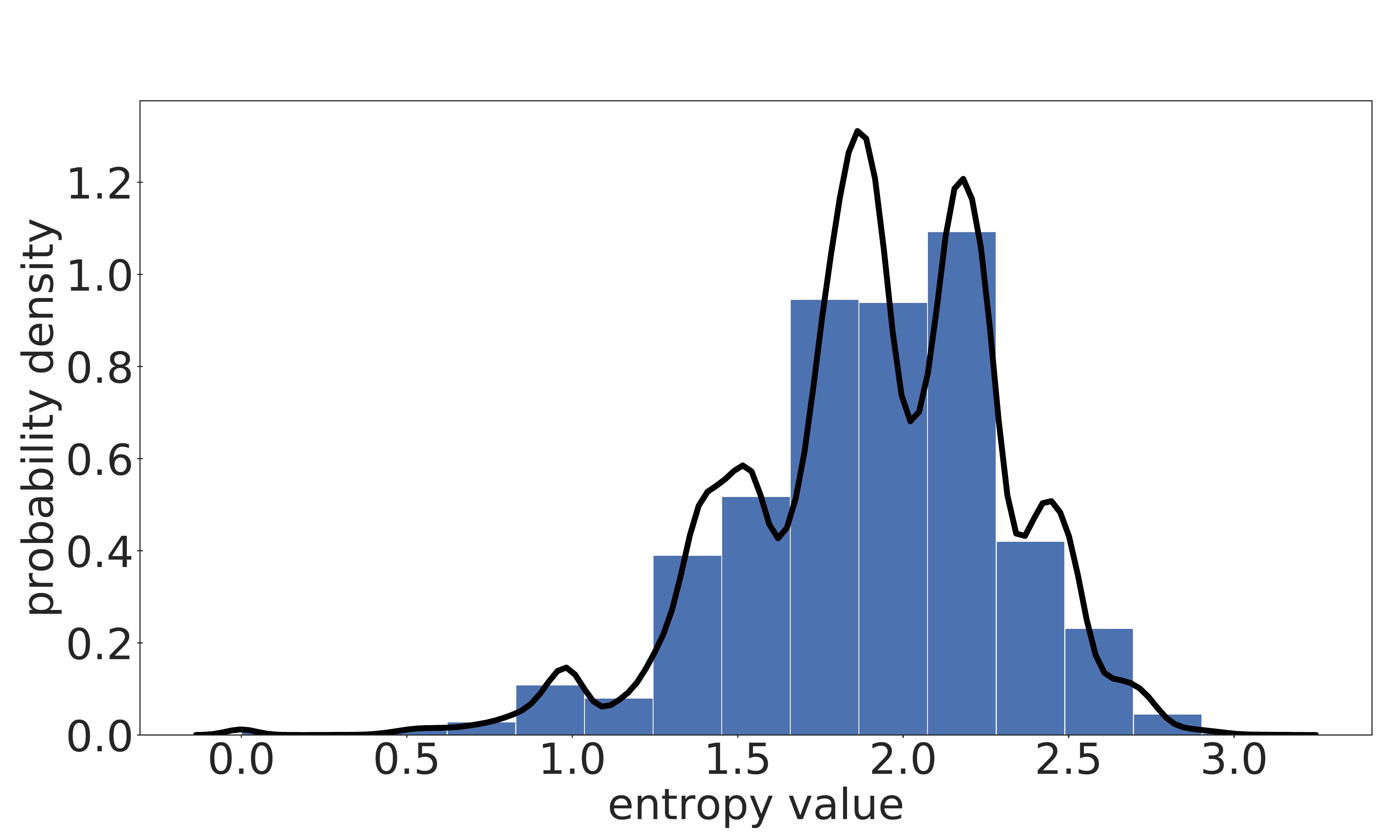}
        \vspace{-20px}
        \caption{Histogram of users' CWTP entropy value. High entropy value means users consider price differently in distinct categories.}
        \label{Figure:entropy}
    \end{minipage}
    \vspace{-10px}
\end{figure}

In this section, we conduct statistical analyses on a real-world dataset, which is collected from one of the biggest e-commerce websites in China (details in Section~\ref{Sec:Exp}).
As stated in Section \ref{Sec:Intro}, price sensitivity depends largely on product category.
To understand the inconsistent sensitivity across categories, we extend the widely used \emph{willing to pay} (WTP) to \emph{category willing to pay} (CWTP) which measures such inconsistency. 
As an indicator that reflects users' price awareness, WTP is defined as the highest acceptable price of an item at which a user is willing to pay~\cite{horowitz2002review}. 
\davy{One step forward, we define CWTP as the highest price a given user is willing to pay for items of a given category. Therefore, for a user who interacted with items of multiple categories, she will have multiple CWTP values (one for a category).
We then compute the entropy of CWTPs for the user, where a small entropy value implies that the user's price sensitivity is consistent across categories, while a large value means that the user considers price differently for products of distinct categories\footnote{For a user u, the entropy value is in the range of [0, $\log C_u$], where $C_u$ denotes the number of product categories that $u$ has interacted with.}. 
Figure \ref{Figure:entropy} plots the histogram of the entropy value over all users. The skewed distribution verifies the aforementioned challenge that the price awareness is highly relevant to the product category, and the inconsistency of price sensitivity across categories exists widely.}
\begin{figure}[t]
    \begin{minipage}[t]{1.0\linewidth}
        \includegraphics[width=\linewidth,bb=0 0 2160 1080]{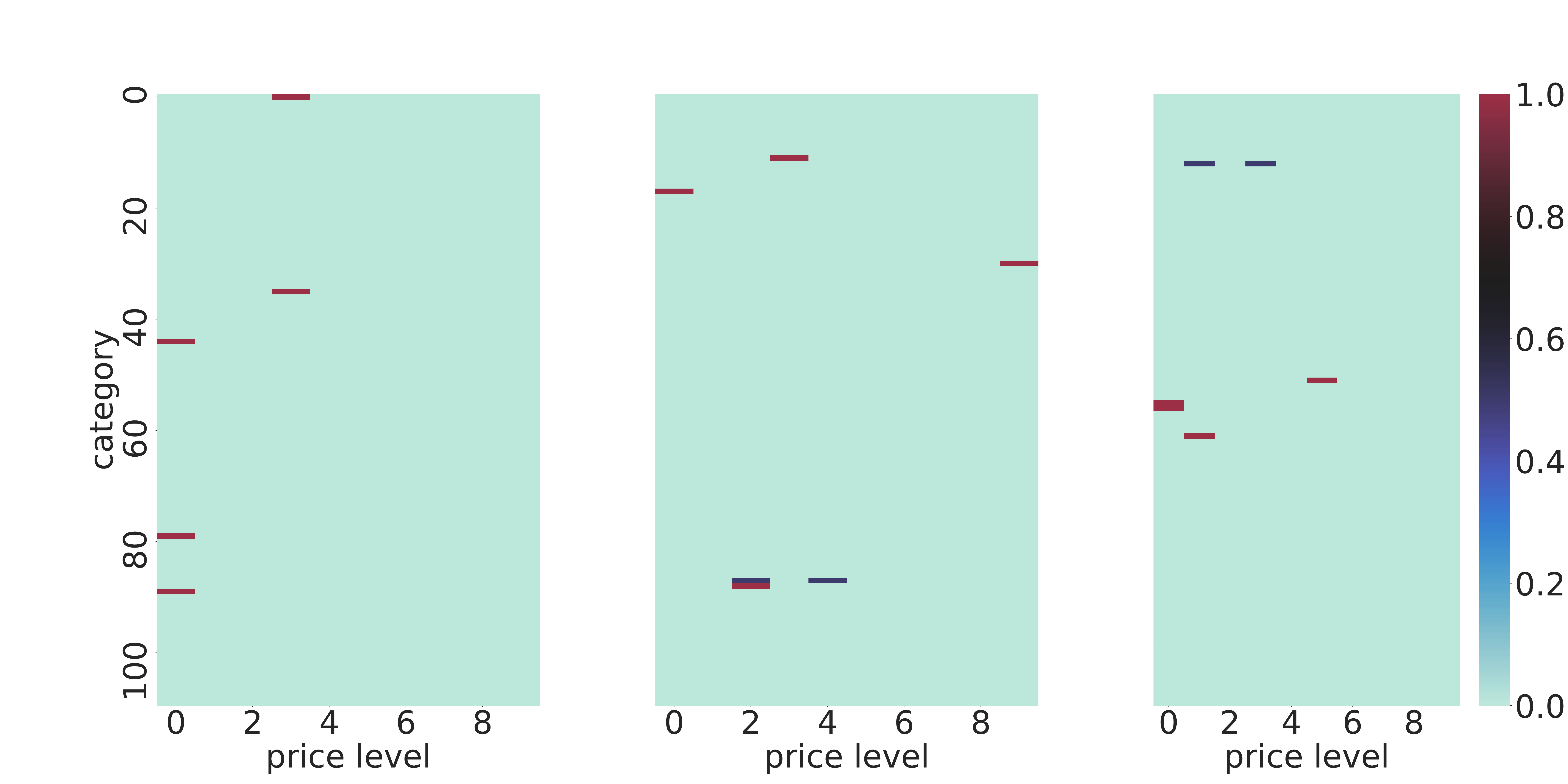}
        \vspace{-20px}
        \caption{Price-category purchase heatmap of three randomly selected users}
        \label{Figure:Pre}
    \end{minipage}
    \vspace{-15px}
\end{figure}

In addition, we randomly sample three users from the dataset, and present the interaction history as a price-category heatmap in Figure~\ref{Figure:Pre}. \davy{We discretize the price using uniform quantization which will be explained in detail in Section \ref{Sec:ProbDef}. In total, there are 110 categories and 10 price levels in this dataset. A row in the price-category heatmap represents one category and a column means one price level. } The heatmap shows that the consumption of a user on a category mostly concentrates on one price level. This implies that the price sensitivity of a user is closely related to product categories. Furthermore, in alignment with the entropy distribution in Fig~\ref{Figure:entropy}, a user is likely to purchase cheap items in one category, but purchases products of a much higher price in another category.

In summary, \davy{through both macro-level statistical analyses and micro-level case studies}, we find that the effect of price is relevant to category and such sensitivity is often inconsistent across different categories. \davy{The observations verify the critical role of category-dependent price awareness --- it is the main difficulty when modeling price in recommendation tasks.}

%% file: 3.probdef.tex
\subsection{Problem Definition}\label{Sec:ProbDef}

The focus of this work is to leverage item price to improve recommendation accuracy.
As discussed above, since the price awareness of a user is closely relevant to product category, it is essential to take category into consideration when designing a system of price-aware recommendation. We formulate this recommendation task as follows.

Let $U$ and $I$ denote the sets of users and items, and $R_{M \times N}$ denote the utility (user-item interaction) matrix where $M$ and $N$ are the number of users and items.  
Here, an observed interaction $R_{ui}=1$ in $R$ means user $u$ once purchased item $i$.
We use $\mathbf{p}=\{p_1, p_2, ..., p_N\}$ and $\mathbf{c}=\{c_1, c_2, ..., c_N\}$ to denote the price and category of items. 

For ease of modeling, we consider price as a categorical variable, discretizing a price value into separate levels using uniform quantization\footnote{In the following of the paper, we will use ``price'' and ``price level'' interchangeably to denote the categorical variable.}.
For example, suppose the price range of category \emph{mobile phone} is [200, 3000] and we discretize it to 10 price levels. 
Then a mobile phone at the price of 1000 will have the price level $\left \lfloor \frac{1000-200}{3000-200}\times 10 \right \rfloor=2$.

Finally, we formulate the problem of price-aware item recommendation as follows:

\textbf{Input:} interaction matrix $R$, price of items $\mathbf{p}$ and category of items $\mathbf{c}$.

\textbf{Output:} The estimated probability of purchasing behavior given a user-item pair $\left(u, i\right)$.

%% file: 4.method.tex
\section{Method}\label{Sec:Method}

\begin{figure*}[t]
    \begin{minipage}[t]{1.0\linewidth}
        \includegraphics[width=\linewidth,bb=0 0 960 200]{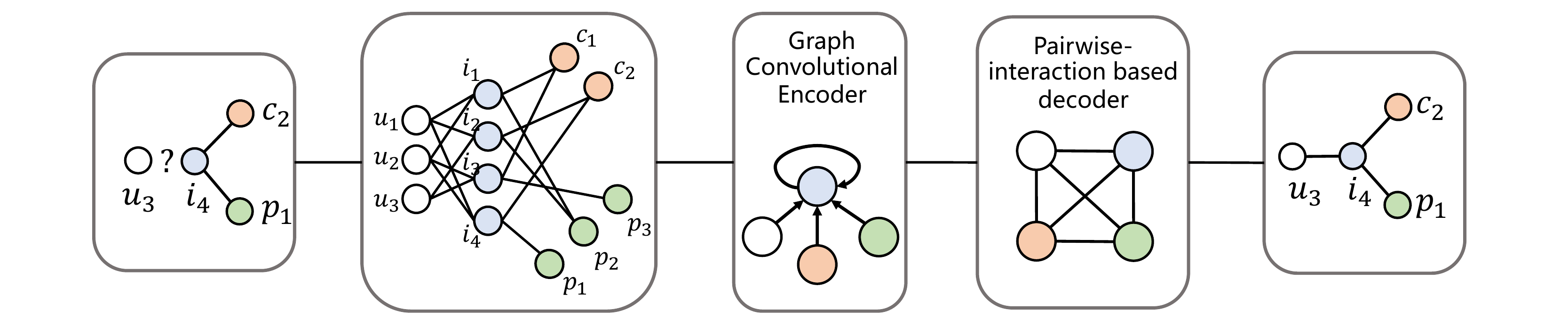}
        \vspace{-10px}
        \caption{The overall design of our proposed PUP method which consists of an unified heterogeneous graph, a graph convolutional encoder and a pairwise-interaction based decoder. The constructed unified heterogeneous graph is composed of four types of nodes where user nodes connect to item nodes, and item nodes connect to price nodes and category nodes}
        \label{Figure:pup}
    \end{minipage}
    \vspace{-5pt}
\end{figure*}
Figure \ref{Figure:pup} illustrates our proposed PUP model. Given a user-item pair $(u, i)$ and the item's two attributes $<p_i, c_i>$ as the input, the model aims to predict the likelihood that $u$ will consume item $i$. Our proposed PUP method is featured with the following three special designs.

\begin{itemize}[leftmargin=*]
    \item Unified heterogeneous graph. 
    To explicitly model user behaviors and item attributes, we discretize the price variable and build a heterogeneous graph with four types of nodes. To tackle the problem of unstated price awareness, we explicitly introduce price as price nodes on the graph instead of input features of item nodes. As for the difficulty of category-dependent influence, we further add category nodes to the graph.
    \item Graph convolutional encoder. To capture both the CF effect and price awareness, we utilize a graph convolutional network as an encoder to learn semantic representations for users, items, prices and categories. With embeddings propagating on the heterogeneous graph, users' price sensitivity is captured by aggregating price-aware information into user nodes.
    \item Pairwise-interaction based decoder. 
    Since the heterogeneous graph consists of four types of nodes which are factorized to a shared latent space, inspired by the philosophy of Factorization Machines \cite{FM}, we employ a pairwise interaction-based decoder to estimate interaction probability.
\end{itemize}
To capture the complicated price factor in recommendation, we estimate the category-dependent and price-aware user preference using two branches --- one branch focuses on a user's interest and models price as a global effect representing a user's overall purchasing power which is unrelated to category, while the other branch focuses on the category-dependent influence of price factor. In this paper, we will call the first branch as the global branch and the second as the category branch. For each branch, we construct a heterogeneous graph and employ a graph convolutional encoder and a pairwise-interaction decoder.
For simplicity, we introduce our method in a single branch manner where there exists only one graph encoder. The two-branch version is similar like mirror image.

\subsection{Unified Heterogeneous Graph}
For the task of price-aware item recommendation, where we have both user-item interaction data and items' price attributes, it is challenging to explicitly capture users' price awareness since user is not directly related to price. In other words, a user's relation with price is build upon the transitive relation of user-to-item and item-to-price. In this way, items play a \textit{bridge} role connecting users and prices.
To address this challenge in capturing the complicated relations into a unified model, we discretize the price variable and build a heterogeneous graph consisting of four types of nodes -- users, items, prices and categories.

Formally, the input interaction data and attributes (category and price) can be represented by an undirected graph $G=\left(V, E\right)$. The nodes in $V$ consist of user nodes $u \in U$, item nodes $i \in I$, category nodes $c \in \mathbf{c}$ and price nodes $p \in \mathbf{p}$. The edges in $E$ are composed of interaction edges $\left(u, i\right)$ with $R_{ui}=1$, category edges $\left(i, \mathbf{c}_i\right)$ and price edges $\left(i, \mathbf{p}_i\right)$ with $i \in I$. The second block in Figure \ref{Figure:pup} illustrates our constructed graph. By introducing four types of nodes, we represent all the entities, features and relations in a unified graph, so as to capture all pairwise relations in an explicit manner.

Notice that we use separate node types for category and price instead of a single node type for the cross feature of (category, price) to avoid redundant parameters. Intuitively, items of the same category with different price shares functionality similarity. Meanwhile, items of the same price from various categories reflect similar price awareness as well. Thus a single type of cross feature lacks connections of the two situations above. By applying distinct node types to category and price, different levels of semantic similarities are captured in the graph.

With respect to graph convolutional networks in the field of node classification~\cite{kipf2016semi}, it is common to use some certain high-level feature vectors like word embeddings extracted by word2vec~\cite{mikolov2013distributed} as the input features of the nodes. Following the same fashion, it seems reasonable to encode price and category information into the input features for user nodes and item nodes which leads to a rather concise bipartite design. However, in our work, we explicitly squeeze out the two important attributes (price and category) as entity nodes to capture the category-dependent price awareness in a more expressive way. Thanks to the explicit design of a heterogeneous graph with four types of nodes which is also in line with the philosophy of Factorization Machines, we could model the pairwise interactions between any features intuitively and effectively. 

As prices and categories are captured directly and explicitly by assigning separate nodes to them, the two aforementioned difficulties of price-aware item recommendation are alleviated. Specifically, the unstated price awareness is transformed to high-order neighbor proximity on the heterogeneous graph which could be well captured by graph convolutional networks. And the category-dependent influence issue is alleviated by linking item nodes to both price nodes and category nodes.

\subsection{Graph Convolutional Encoder}
Latent factor model (LFM), which tries to encode entities in a low-dimensional latent space, is a widely-used mechanism in recommender systems~\cite{hu2008collaborative, mnih2008probabilistic}. For traditional LFMs in recommender system, such as Matrix Factorization, an observed $(u, i)$ pair will push $u$ and $i$ to each other in latent space. However, in our built unified heterogeneous graph, there are two more pairs, $(i, p)$ and $(i, c)$ . Besides, there are underlying user-price interaction $(u, p)$ when user $u$ purchases item $i$ with price $p$. In this paper, we extend traditional LFMs that only learn representations for users and items, and try to learn representations of four types of entities in the same latent space. Recent research~\cite{kipf2016semi,GraphSAGE,GC-MC} have shown that performing message passing on the graph could lead to semantic and robust node representations for multiple tasks like node classification and link prediction. A special class of algorithms among them called Graph Neural Networks achieve the state of art in the field of network representation learning.
We employ an encoding module consisting of an embedding layer for converting one-hot input to low-dimensional vectors, an embedding propagation layer to capture both CF effect and price awareness, and a neighbor aggregation layer to model neighbor similarity.

\textbf{Embedding Layer.} 
As discussed previously, in our proposed model, since price attributes and category attributes are squeezed out as nodes, ID is the only feature left for a node. Thus, we introduce an embedding layer to compress the one-hot ID encoding to a dense real-value vector. That is, we represent each node with a separate embedding $e' \in \mathbb{R}^d$, where $d$ is the embedding size.

\textbf{Embedding propagation layer}
In GCN, embeddings of nodes propagate to their first order neighbors and further if more than one convolutional layer are applied. In our encoder, the embedding propogation layer captures the message transferred between two directly connected nodes which could be user-item, item-price or item-category. Suppose node $i$ and node $j$ are two connected nodes in our unified heterogeneous graph. The propagated embedding from node $j$ to node $i$ is formulated as follows:
\begin{equation}
    \mathbf{t}_{ji}=\frac{1}{\left|\mathcal{N}_i\right|}e_j',
\label{Eq:propogate}
\end{equation}
where $\mathcal{N}_i$ denotes the set of neighbors for node $i$ and $e_j'$ is the embedding of node $j$ retrieved from the embedding layer. As stated in \cite{SGC}, adding self-loops is of significant importance to GCN since it shrinks the spectrum of the normalized Laplacian, thus we link each node in our heterogeneous graph to itself which makes node $i$ also appear in $\mathcal{N}_i$. 

\textbf{Neighbor aggregation layer.} 
From the perspective of network representation learning, the neighbor relation of two nodes in the graph structure implies that their representations should also be near in the transformed latent space. Inspired by recent advances in graph convolutional networks~\cite{GC-MC, GraphSAGE,PinSage}, we update the representation of a node by aggregating its neighbors' representations. Among all the aggregating operations, summation, taking average, and LSTM are the most frequently used approaches~\cite{GraphSAGE,xu2018powerful}. In our proposed encoder, we adopt average pooling and utilized a non-linear activation function to perform message passing on the graph.

Specifically, let $e_u$, $e_i$, $e_c$, and $e_p$ denote the representations for user $u$, item $i$, category $c$ and price $p$. The updating rule can be formulated as follows:
\begin{equation}
\begin{aligned}
    o_u = & ~\sum_{j\in\{i~\text{with}~R_{ui}=1\}\cup \{u\}}\mathbf{t}_{ju},&\\
    o_i = & ~\sum_{j\in\{u~\text{with}~R_{ui}=1\}\cup \{i,\mathbf{c}_i,\mathbf{p}_i\}}\mathbf{t}_{ji},&\\
    o_c = & ~\sum_{j\in\{i~\text{with}~\mathbf{c}_i=c\}\cup \{c\}}\mathbf{t}_{jc},&\\
    o_p = & ~\sum_{j\in\{i~\text{with}~\mathbf{p}_i=c\}\cup \{p\}}\mathbf{t}_{jp},&\\
    e_f = &~\tanh(o_f), f \in \{u,i,c,p\}.\\
\end{aligned}
\end{equation}

\begin{figure}[t]
    \begin{minipage}[t]{\linewidth}
        \includegraphics[width=\linewidth,bb=0 0 960 540]{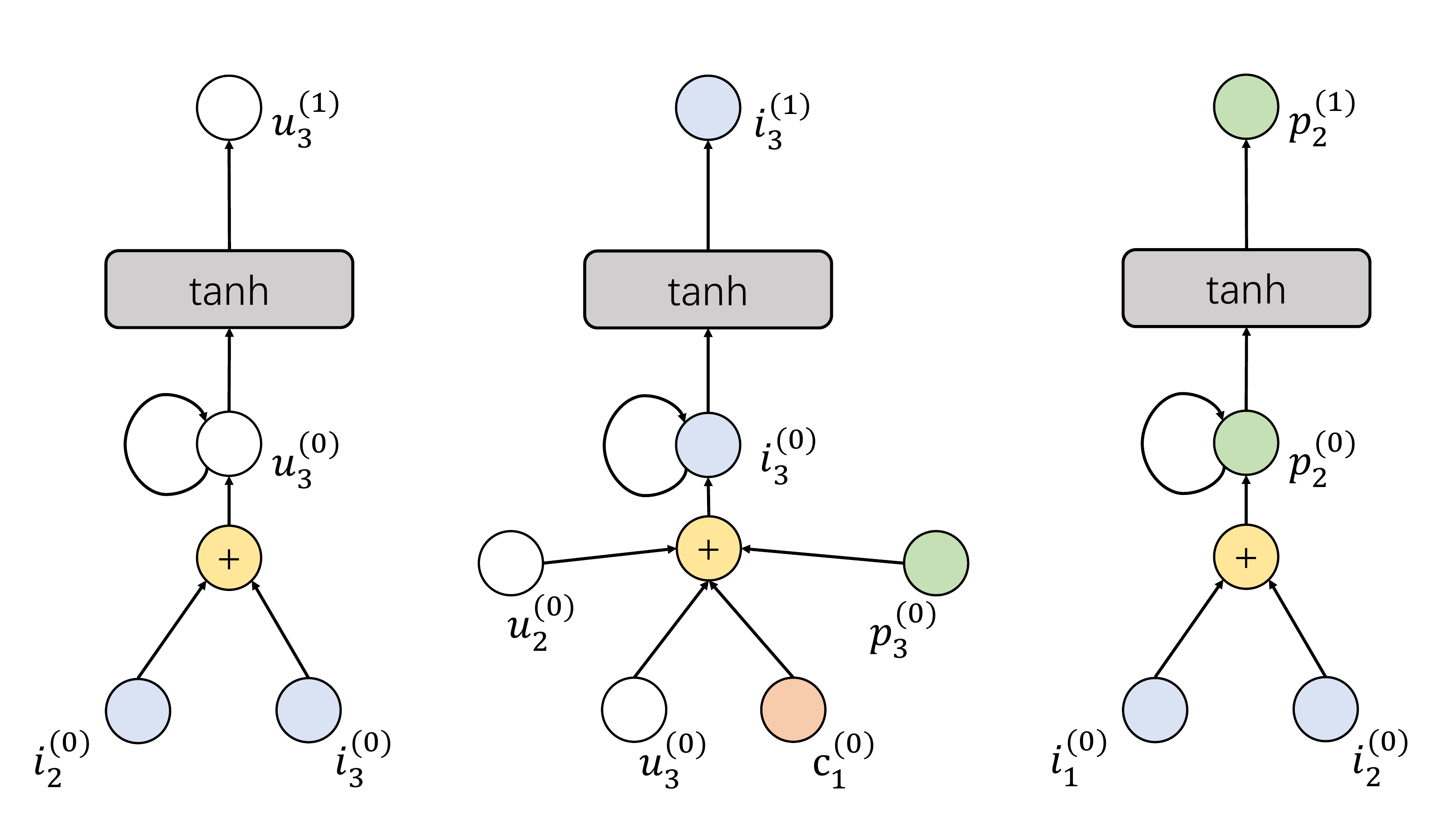}
        \caption{The updating rule for different types of nodes. Left: user node. Mid: item node. Right: a price node. Category nodes update similarly with price nodes}
        \label{Figure:update_rule}
    \end{minipage}%
\end{figure}

Figure~\ref{Figure:update_rule} illustrates the updating rule for different types of nodes. Due to the intrinsic expressive power of embedding propagation and neighbor aggregation, the learned representations extracted by the graph convolutional encoder can effectively model the relations between nodes and their high order neighbors. 

Intuitively, items of the same price level are likely to be more similar than items of different price levels. In our constructed heterogeneous graph, a price node links to all the items of that price level, and the graph convolutional encoder guarantees that the output representations for those items will absorb the price embedding into themselves by embedding  propagation and neighbor aggregation. Therefore, the encoder generates item representations with price-aware similarities. Category-aware similarities are captured in the same way since a category node is connected to all the item nodes which belong to that category.

Furthermore, a user's price awareness is largely reflected by her interacted items and other users' purchase history through collaborate filtering. Thus it is crucial to leverage items as the bridge between users and price awareness. In our model, a user's representation is aggregated from her interacted items explicitly and the items are directly linked to categories and prices. Thus category nodes and price nodes are high-order neighbors with respect to user nodes, and the price awareness is propagated to the users via intermediate item nodes.

From the perspective of recommendation, the proposed graph convolutional encoder is able to capture the similarity of any two nodes when there exists a path between them. With respect to the classic Matrix Factorization algorithm, collaborative filtering effect is captured implicitly by optimizing to estimate user-item interactions. However, in our graph convolutional encoder, we explicitly incorporate collaborate filtering effect by aggregating a node's neighbors. Specifically, similar users who have interacted with the same item are 2-order neighbors on the heterogeneous graph and this proximity could be captured by the graph convolutional encoder.

\subsection{Pairwise-interaction Based Decoder}
As discussed previously, we adopt a two-branch design to estimate user-item interactions with emphasis on incorporating price into recommendation. The global branch models the price effect in a large scope which focuses on a user's overall purchasing power. The category branch instead concentrates on a rather ``local" level where the category factor influences a user's price sensitivity. For each branch, we employ a pairwise-interaction based decoder to estimate the interaction probability and combine the two predicted scores as the final result.

As we represent users, items, categories and prices as four types of nodes on a unified heterogeneous graph, the learned representations for different types of nodes share the same latent space. Inspired by Factorization Machines~\cite{FM} which factorizes all the features in a shared latent space and estimates the interaction by taking inner products of every pair of feature vectors, we employ a decoder following the FM fashion. Formally, using the same notation from the previous encoder section, the estimated purchase probability between user $u$ and item $i$ of category $c$ and price $p$ can be formulated as follows:
\begin{equation}
    \begin{aligned}
        s &= s_\mathrm{global} + \alpha s_\mathrm{category} \\
        s_\mathrm{global} &= e_u^Te_i + e_u^Te_p + e_i^Te_p \\
        s_\mathrm{category} &= e_u^Te_c + e_u^Te_p + e_c^Te_p,
    \end{aligned}
\end{equation}

where the final prediction combines the results from two branches with a hyper-parameter $\alpha$ to balance the two terms. It should be noted that the each branch has its own graph convolutional encoder, thus the embeddings used for computing $s_\mathrm{global}$ and $s_\mathrm{category}$ are different and independent.

With respect to the global branch, three features which are users, items and prices are fed into a pairwise-interaction based decoder in a 2-way FM manner. In this branch, the three inner products each captures the user's interest, the user's global price effect and the item's price bias respectively. we estimate the interaction probability without category embeddings and thus category nodes only serve as a regularization term on the graph which makes items of the same category near to each other. Since category information is hidden in the decoder process in the global branch, the \textit{local} effect of price which is related to category is pushed out from the learned latent space. And the global price influence, which reflects a user's overall purchasing power and affordability, is reserved in the latent space learned by the powerful graph convolutional encoder.

However, as discussed previously, users' price sensitivity is largely relevent to category and often appears inconsistently across different categories. Thus we add a category branch which serves to capture this subtle price awareness which is related to category. In this branch, we omit item embeddings when estimating interactions and only take users, categories and prices into consideration. Item nodes simply play the \textit{bridge} role transferring price and category information to users. By taking inner products of the three embeddings, users' category-dependent price awareness is guaranteed in the shared latent space.

Following the graph convolutional encoder with strong power in learning representations, we employ a two-branch pairwise interaction based decoder to estimate the interaction probability. The design of this decoder largely benefits from the philosophy of Factorization Machines\cite{FM} which is one of the most widely used and effective method for estimating CTR (click through rate) in both research and industry. Moreover, our decoder serves great interpretability with respect to price awareness since the two-branch design disentangles the global effect and the category-dependent effect of the price factor in recommender systems.

\subsection{Model training}

\textbf{Semi-supervised graph auto-encoder.} To train our proposed PUP model, we follow a popular fashion of semi-supervised graph auto-encoder~\cite{GC-MC, tu2018deep, bojchevski2018deep}.
That is, during the encoding stage, we utilize a GCN which aims at learning expressive and robust representations for all the four types of nodes. While in the decoding stage, we only focus on reconstructing user-item edges on the heterogeneous graph and omit item-price and item-category edges, because predicting user-item interaction is the main task for recommendation.

\textbf{Loss function.} In order to learn users' preferences on different items, we adopt Bayesian Personalized Ranking (BPR)\cite{BPR} as our loss function which has been widely used in recommendation tasks for implicit data~\cite{pasricha2018translation, zhang2016collaborative, ncf}. The BPR loss induces the model to rank the positive samples (interacted items) higher than negative samples (unobserved interactions). This pairwise object which focuses on the relative preference priority of items instead of absolute interests could be formulated as follows:
\begin{equation}
    L = \sum_{\left(u, i, j\right) \in \mathcal{O}}-\ln\left(\sigma\left(s\left(u,i\right)\right)-\sigma\left(s\left(u,j\right)\right)\right) + \lambda\left\|\Theta\right\|^2,
\label{Eq::loss}
\end{equation}
where $\mathcal{O}$ denotes the set of positive-negative sample pairs and $\sigma$ stands for \textit{sigmoid} function. The second term of equation (\ref{Eq::loss}) performs L2 regularization where $\Theta$ stands for model parameters and $\lambda$ controls the penalty strength.

%% file: 7.implementation.tex
\section{Implementation}\label{Sec:Implementation}
In this section, we introduce several important details when implementing our proposed PUP model.
\subsection{How to Perform Graph Convolution?}
As introduced previously, our proposed graph convolutional encoder is composed of an embeddding layer, an embedding propagation layer and a neighbor aggregation layer. Message passing on the heterogeneous graph is accomplished through stacking the three layers. In fact, similar to the semi-supervised approach proposed in \cite{kipf2016semi}, the graph convolutional encoder can be implemented effectively utilizing sparse matrix production.

We define the rectified adjacency matrix $\hat{A}$ which is sparse as follows:
\begin{equation}
    \hat{A} = f\left(A + M_I\right),
\end{equation}
where $A$ is the original adjacency matrix and $M_I$ is the identity matrix. $f\left(M\right)$ takes average on each row of matrix $M$. Adding the identity matrix is equivalent to adding self-loops on the graphs which will influence the performance of GCN significantly~\cite{SGC}.

We denote the input one-hot encoding feature matrix as $F_{\mathrm{in}}$ where each row represents a node, and use $W$ to represent the learnable embedding matrix. Then the output representation matrix is computed as follows:
\begin{equation}
    F_{\mathrm{out}} = ~\tanh(\hat{A}F_{\mathrm{in}}W).
\end{equation}
It can be easily proved that the matrix production expression above is equivalent to the design of stacking an embedding layer, an embedding propagation layer and a neighbor aggregation layer.

\subsection{How to Efficiently Perform Decoding?}
Since in the pairwise-interaction based decoder we take inner products of every pair of features, the computation complexity is relatively high. However, the computation can be reduced to linear complexity using the following trick which was first introduced in \cite{FM} (take the category branch as an example):
\begin{equation}
    \sum_{f,g \in \left\{u,c,p\right\}, f \neq g}{e_f^Te_g} = \frac{1}{2}\left(\left(\sum_{f}e_f\right)^2 - \sum_{g}\left(e_g^2\right)\right).
\end{equation}
It is worthwhile to state that more features could be added to our model and the computation complexity will matter a lot when there exist hundreds of features which is quite common in today's recommender systems.

\subsection{How to Prevent the Model from Overfitting?}
Dropout is an effective way to prevent neural models from overfitting~\cite{dropout}. We adopt dropout on the feature level, which means randomly dropping the output representations with probability $p$. $p$ is a hyper-parameter in our method. We perform grid search on this hyper-parameter and report the best performance. With the help of dropout technique, our proposed PUP method learned more robust node representations on the unified heterogeneous graph.

%% file: 5.experiments.1.tex
\section{Experiments}\label{Sec:Exp}

In this section, we first investigate the performance of our proposed PUP method compared with existing baselines and verify the effect of price-aware recommendation system. One step forward, we dive deeper into the role price factor plays in our proposed method. Then we study the effect of our two-branch design to see whether price awareness is carefully modelled in our method. Furthermore, we test the performance on users with different consistency in terms of the price awareness across categories. Moreover, as stated in~\cite{SIGIR14}, incorporating the price factor could improve the performance when recommending items of unexplored categories which is a cold-start problem. Thus we also measure our proposed model under this protocol.
		
\subsection{Experimental Settings}
\subsubsection{Dataset and Evaluation Protocol}
To evaluate the performance of our proposed PUP method, we utilize two real-world datasets for comparison: Yelp and Beibei, which both have abundant category and price information for items. Statistics of these two datasets are summarized in Table \ref{tab::dataset}.

\input{tables/dataset.tex}
\begin{itemize}[leftmargin=*]
\item  \textbf{Yelp}. We adopt Yelp2018 Open Dataset\footnote{https://www.yelp.com/dataset} in which restaurants and shopping malls are regarded as items. We choose all the sub-categories under the top-level category \textit{restaurant}. \davy{In this dataset, price of each restaurant is shown as different number of dollar symbols which ranges from 1 to 4. Thus we directly use the number of dollar symbols as price levels in our experiments.} Finally, we utilize 10-core settings which means only retaining users and items with at least 10 interactions.
\item \textbf{Beibei}. This is a dataset collected from one of the largest E-commerce platforms\footnote{https://www.beibei.com} in China. In this dataset, all items are with specific category and price information. \davy{Since the price of each item in this dataset is of continuous form, we discretize the continuous price to 10 price levels using uniform quantization and use the 10-core settings to guarantee data quality.}
\end{itemize}

For each dataset, we first rank the records according to timestamps and then select the early 60\% as the training set,  middle 20\% as the validation set, and the last 20\% as the test set. 
For each user, the items that are not interacted by the user are viewed as negative samples. We perform negative sampling to constitute positive-negative sample pairs for training. To evaluate the effectiveness of top-K recommendation, we use the same metrics as in~\cite{ncf} including Recall and NDCG. We report average metrics for all the users in the test set.

\subsubsection{Baselines}
To show the effectiveness of our proposed PUP method, we compare the performance with the following baselines.
\begin{itemize}[leftmargin=*]
   \item \textbf{ItemPop} It is a non-personalized method that ranks items just according to their popularity in the training set.
   \item \textbf{BPR-MF~\cite{BPR}} This is a standard matrix factorization model optimized by Bayesian Personalized Ranking (BPR)~\cite{BPR} loss.
   \item \textbf{PaDQ~\cite{SIGIR14}} This method is based on CMF~\cite{CMF} to handle price information. Specifically, it factorizes user-item, user-price and item-price matrix simultaneously with shared latent representations among matrices.
   \item \textbf{FM~\cite{FM}} Factorization Machines (FM) is a competitive model which applies a sum of pairwise inner product of user or item features to obtain the prediction score. In our experiments, we integrate price and category into FM by regarding them as item features.
   \item \textbf{DeepFM~\cite{DeepFM}} This method is an ensemble model that combines Factorization Machines and deep neural networks to capture both low- and high- order feature interactions. We treat price and category similarly as in FM.
    \item \textbf{GC-MC~\cite{GC-MC}} It adopts graph convolutional networks to learn the representations of users and items on a bipartite user-item graph. We use one-hot ID features as input features for user and item nodes.
    \item \textbf{NGCF~\cite{NGCF19}} This method is a new recommendation framework based on graph neural networks which explicitly captures the collaborative signal by performing embedding propagation on a bipartite user-item graph. In our experiments, we use a concatenation of one-hot ID feature and one-hot price feature as the input feature for item nodes.
\end{itemize}
We did not compare our method with FMF proposed by~\cite{WWW17} and $\mathrm{SVD_{\textit{util}}}$ proposed by \cite{wang2011utilizing} since these methods incorporate the \textit{dynamic} of price or net utility which we leave it for future work.

\subsubsection{Parameter settings}
\input{tables/performance_comparison_ICDE_revision.tex}
We adopt BPR loss for all methods and fix the embedding size as 64. We use Adam for optimization with the initial learning rate as 1e-2. The size of mini-batch is fixed at 1024 and negative sampling rate is set to 1. For each model, we train for 200 epochs and reduce the learning rate by a factor of 10 twice for convergence.

\subsection{Performance Comparison}
We first compare the results of all the methods on two datasets with respect to: Recall@50, NDCG@50, Recall@100 and NDCG@100.
Table \ref{tab::performance_comparison} presents the overall comparison of our proposed PUP method and other baselines. From the results, we have several important observations.

\subsubsection{Incorporating Price into Recommendation Improves the Accuracy}
Generally, incorporating more attributes and features into recommender systems would increase the overall recommendation performance. Compared to trivial CF methods like MF which only consider users and items, attribute-aware methods are able to capture the relationship between much more features and interactions. In this work, we only focus on two attributes which are category and price. As illustrated in the results, attribute-aware methods outperform other trivial CF methods. With respect to price-awareness modeling, experimental results in both datasets verify that incorporating price into recommendation could attain improvements in accuracy. Specifically, FM and DeepFM outperform BPR-MF, and NGCF outperforms GC-MC in most cases. As stated previously, the input feature for NGCF contains price information while the input of GC-MC is just one-hot ID encoding. However, the performance gain is not significant, which indicates that the two challenges of price-aware recommendation remain unresolved in these attribute-aware methods.

\subsubsection{Price Should be Considered More as an Input Rather Than a Target}
PaDQ and FM, which are two typical methods of attribute-aware recommendations, differ in how they incorporate price into the system. Specifically, PaDQ works in a generative way, which means it takes price into consideration by adding extra tasks of predicting price awareness with shared underlying latent factors. However, FM follows a deterministic fashion which means regarding user or item IDs as special attributes and factorizes all attributes to the same latent space to make predictions. As the results illustrated, FM substantially surpasses PaDQ and PaDQ is even worse than BPR-MF. The large performance gap indicates that price should be considered more as an input of recommendation rather than a target to predict.

\subsubsection{Neural Based Methods and Graph Based Methods Have an Advantage over Other Methods}
From the results, neural based methods and graph based methods achieves better results than other ``shallow" models in most cases. This performance gain is reasonable since neural units increase the capacity of the model and the graph structure is much more expressive than a look-up embedding table. Specifically, DeepFM, GC-MC and NGCF generally attains better results. DeepFM combines DNN and FM to capture low- and high- order interactions simultaneously. GC-MC applies graph convolution on the user-item bipartite to learn more meaningful representations for users and items. NGCF captures the CF effect by performing embedding propagation on the graph. The results show that it is promising to enhance representation learning and interaction modeling in recommendation by introducing neural networks and graph neural networks into the system.

\subsubsection{Our Proposed PUP Method Achieves the Best Performance}
Our proposed PUP consistently achieves the best results on all metrics of both datasets. Following the semi-supervised graph auto-encoder fashion, this two-branch method which is specifically designed for modeling price awareness is proven to be effective. Reuslts of t-tests indicate that the improvements are statistically significant for $p < 0.005$.

To summarize, the extensive comparisons on two datasets verify that our proposed PUP method is able to effectively leverage price of items to improve recommendation.

\subsection{The Effect of Price Factor}
In this section, we perform ablation study to verify the importance of incorporating price into recommender systems. Several modified models with the price factor removed are constructed for comparison with our proposed PUP method. Furthermore, in modern e-commerce platforms, the price range is often large and the distribution of price tends to be much complicated. The uniform quantization process of price factor might fail to capture the price preference when price is not uniformly distributed. Therefore, we adapt the quantization process according to the rank of price and compare the two quantization methods. With respect to the discretization process of price factor, the number of price levels has a great influence on the fineness of capturing the price preference which directly impacts the performance of the recommendation system. Thus we conducted experiments on different price levels to investigate how the granularity of price affects the recommendation results. 
To confirm the importance of the price factor, we conducted experiments on a general dataset which is collected from the reviews on Amazon\cite{he2016ups} and the original price information is available. We selected product reviews of 5 categories (\textit{Cell Phones and Accessories}, \textit{Tools and Home Improvements}, \textit{Toys and Games}, \textit{Video Games} and \textit{Beauty}). We utilized the 5-core version provided by \cite{he2016ups} which resulted in a total number of 438355 interactions with 48424 users and 33483 items.

\input{tables/ablation.tex}
\subsubsection{Ablation Study of Price Factor}
In our proposed two-branch design, the price factor is incorporated into recommender systems from both global level reflecting users' purchasing power and local level focusing on category-dependent price awareness. We constructed several slim versions of our PUP model to verify the necessity of taking price into consideration. In these slim versions, category factor, or price factor, or both are removed. Results are illustrated in Table \ref{tab::ablation}. From the results we can address that incorporating price into recommender systems brings significant improvements with respect to recommendation accuracy and the price factor is indeed a crucial feature in e-commerce recommendation scenarios. Nevertheless, jointly modeling the price factor and the category factor achieves the best performance which verify the necessity of taking category into consideration. The large performance gap between PUP and other slim versions confirms that our proposed method is of great effectiveness in capturing users' price awareness.

\input{tables/rank_uniform.tex}

\subsubsection{The Quantization Process of Price Factor}
In most cases, the distribution of price factor on e-commerce platforms is complicated. Usually there are a few products with extremely high or particularly low price which in turn makes the price range large. However, the majority of the products often lie in the middle of the range. The non-uniform distribution of price factor requires subtle adaptation of the uniform quantization process. We adopt rank-based quantization to alleviate the problem of uniform quantization whose performance is suboptimal when the price distribution is complex. In uniform quantization, we calculate the normalized price by subtracting the minimum price and then divided by the price range of the corresponding category. However, in rank-based quantization, we first rank the products by price within their categories and then transform the rank to percentile form. At last, the discretized price is obtained by multiplying the total number of price levels and taking the integer part. Results of the two different quantization processes are shown in Table \ref{tab::rank_uniform}. It is reasonable that rank-based quantization attains much better results than uniform quantization since the price factor is not uniformly distributed. Rank-based quantization transforms the non-uniform distribution of the absolute price value to the uniformly distributed ranking percentage value. This adaptation of the quantization process of price factor makes our proposed PUP model still feasible under the circumstance of large price range and complicated distribution of price factor.

\subsubsection{The Fineness of the Price Factor}
\begin{figure}[t]
    \begin{minipage}[t]{1.0\linewidth}
        \includegraphics[width=\linewidth, bb=0 0 1440 864]{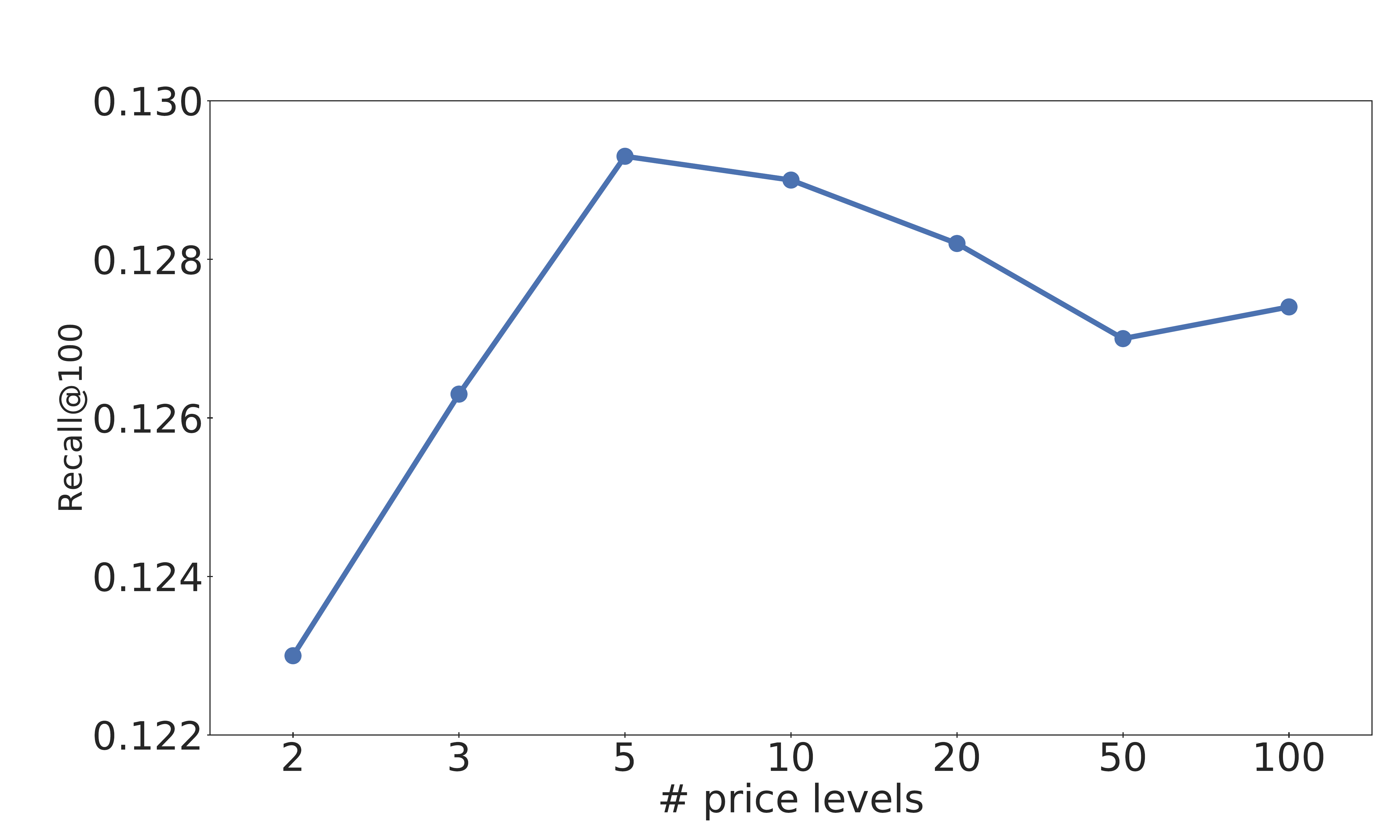}
        \vspace{-20px}
        \caption{Performance on amazon at different number of price levels.}
        \label{Figure::price_level}
    \end{minipage}
    \vspace{-15pt}
\end{figure}

Usually price is of continuous form on e-commerce platforms. However, utilizing the price factor directly as an absolute value brings much complexity to the system since nodes on the heterogeneous graph are discrete in essence. Nevertheless, consumers tend to pay attention to other factors like brand or sales when the candidate items cost roughly the same. Therefore, we translate the price to price levels when it is continuous for both simplicity and rationality. The number of price levels is an important option which decides the fineness of the price factor in our proposed method. We experimented on different price levels to study how the granularity of the price factor influences the recommendation performance. Figure \ref{Figure::price_level} shows the results of different price levels. When the number of price levels is set extremely low such as two which means the model only makes a coarse discrimination between cheap and expensive, the price factor is not accurately incorporated, thus the overall performance is inferior to more elaborate models with finer capture of price factor. While if we set the number of price levels too high like 100, items of near price are allocated to different price levels which could damage the performance as well since the difference in price is not so important under this condition.

\subsection{The Two-branch Design}
Previously we introduced our two-branch design which aims at disentangling the global and local effect of price awareness in recommendation.
If we fix the holistic embedding size, the specific dimension at which we slice the embedding requires careful consideration. With more dimensions allocated to the global branch, the PUP model pays more attention to users' interest and the average influence of the price factor. With the category branch getting more dimensions, the local and category-dependent price awareness plays a more important role in estimating interactions. Intuitively, larger embedding size leads to more capacity and expressive power in that branch.

We slice at different dimension to study how the allocation of embedding size influence price awareness modeling. Results are shown in Table \ref{tab::slicing}. In the table, an allocation of $m/n$ indicates the embedding size is $m$ for the global branch and $n$ for the category branch. It is reasonable that better performance is achieved when the global branch takes the majority since items is of vital importance when estimating user-item interactions while in the category branch item embeddings are hidden. However, further compressing the embedding size of the category branch to extremely low such as 4 or lower will worsen the recommendation accuracy. Extremely low embedding size restricts the capacity of the category branch and thus the category-dependent price awareness could not be well captured which leads to inferior performance.

\input{tables/embedding_slicing.tex}
\input{tables/user_group_ICDE.tex}

\subsection{Consistency of Price Awareness across Categories}
It has been introduced that the effect of price on influencing users' purchase behavior largely depends on item categories. Whether the price awareness is consistent across categories is crucial when making recommendations for items of various categories. We divide users into groups according to their entropy value of CWTPs which was defined in Section \ref{Sec:Pre}. Table~\ref{tab::entropy} shows the performance of our proposed PUP method on different user groups compared to DeepFM. From the results, We have the following two findings:
\begin{itemize}[leftmargin=*]
   \item Both DeepFM and PUP perform much better on consistent users than on inconsistent users. The reason for the performance gap is that it is more difficult to predict users' interest when they regard price differently over distinct categories.
   \item Our proposed PUP achieves better results on both consistent users (about 41.76\% boost) and inconsistent users (about 1.18\% boost). Given the fact that capturing the preference of inconsistent users is more challenging, our proposed PUP method is still able to improve the performance due to its powerful capacity of learning high-quality representations on the constructed unified heterogeneous graph.
\end{itemize}
The inconsistency of price awareness across categories brings much more difficulties when incorporating price into recommendation. The performance gap between DeepFM and PUP suggests that it is not sufficient to model this inconsistency using a unified model and it requires specific design towards this inconsistency. In our two-branch structure, this inconsistency is exactly what the category branch aims to model. By combining global and local effect of the price factor, our proposed PUP method could improve the recommendation performance on users with diverse consistency of price awareness.

\subsection{Utilizing Price to Tackle Cold-Start Problems}
\begin{figure}[t]
    \begin{minipage}[t]{1.0\linewidth}
        \includegraphics[width=\linewidth,bb=0 0 1440 864]{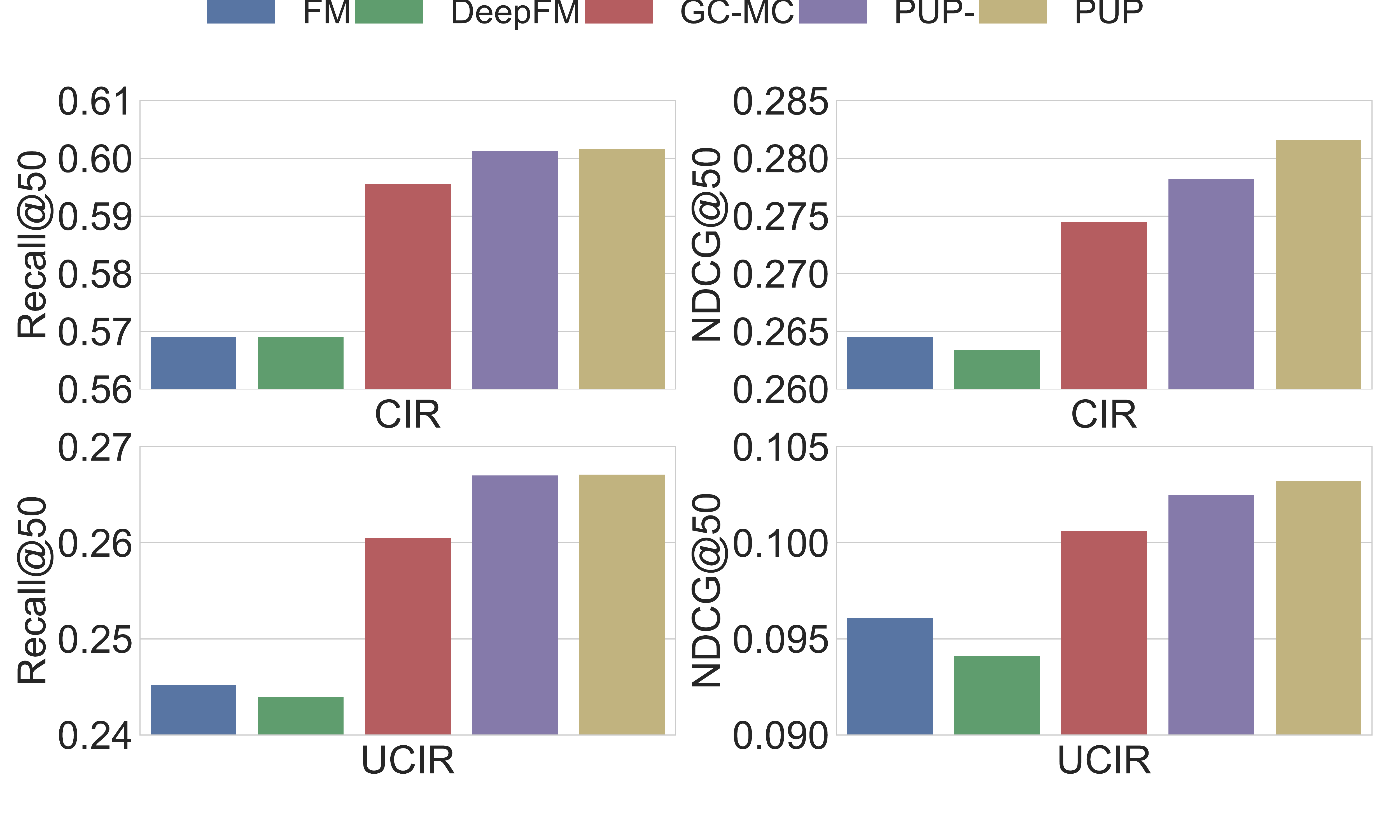}
        \vspace{-30px}
        \caption{Performance comparison on unexplored categories of Yelp dataset}
        \label{Figure:cir_ucir}
    \end{minipage}
    \vspace{-25pt}
\end{figure}
In modern e-commerce platforms, a user usually only interacts with a small number of categories compared with the total number of categories on the platform. However, recommending items of unexplored categories is rather important both for diversity and for maximizing the revenue. This task is a so called cold-start problem since there exists limited data records in the unexplored categories. Users' preference and price awareness are not always consistent across categories. Therefore, what we have learned on explored categories cannot be directly transferred to unexplored ones. \cite{SIGIR14} first proposed that the price could come to help in this situation.

To evaluate the performance on unexplored categories, we make a few adaptations to our datasets. We find those users who purchase items in the test set from categories that are different from those purchased categories in the training set. Then we filter out those items in the test set belonging to explored categories. We conduct experiments according to two protocols which were utilized in~\cite{SIGIR14}:
\begin{itemize}[leftmargin=*]
    \item \textbf{CIR} (Category item recommendation): For this protocol, the candidate item pool is composed of all the items which belongs to the test positive unexplored categories.
    \item \textbf{UCIR} (Unexplored category item recommendation): For this protocol, the candidate item pool consists of all the items which are not in the train positive categories.
\end{itemize}
For example, suppose there are 7 categories in total which are $\left\{A,B,C,D,E,F,G\right\}$. A user purchases items of category $A$, $B$ and $C$ in the training set and purchases items of category $E$ in the test set. Then according to CIR protocol, all the items of category $E$ forms the candidate item  pool. However, in UCIR problem, the candidate item  pool is composed of items from unexplored categories which are $\left\{D,E,F,G\right\}$.

Figure \ref{Figure:cir_ucir} shows the performance comparison on unexplored categories. PUP- stands for a slim version of PUP with category nodes removed. The results verify the positive effect of incorporating price when recommending items of unexplored categories. Specifically, PUP and PUP- outperforms GC-MC in all cases.
Without incorporating price and category, GC-MC only captures user-item interactions. And when performing neighbor aggregation, item nodes of unexplored categories can only be reaches via intermediate user nodes in GC-MC, while in PUP or PUP- price nodes can also be leveraged which makes it much easier to transfer from explored categories to unexplored categories. 

Moreover, the results illustrates that GCN based methods (GC-MC, PUP-, PUP) consistently outperform factorization based methods (FM, DeepFM). The performance gap indicates that a simple look-up embedding table is insufficient to learn high-quality representations under cross category protocols. With the help of GCN, item nodes of unexplored categories could be reached by neighbor aggregation on the graph.

Finally, our proposed PUP model achieves the best performance in both CIR and UCIR problems on two datasets. Due to the powerful graph convolutional encoder, our proposed PUP is able to capture users' preference on unexplored categories. In our contructed unified heterogeneous graph, item nodes linked to unexplored categorie nodes are high order neighbors for the user since those item nodes could be reached via price nodes and user nodes connected to them. With strong collaborative filtering effect, an item of an unexplored category could be reached within 3 hops (user-item-user-item). Moreover, if the user purchases items of enough price levels, the item could also be reached through price nodes (user-item-price-item). 

To summarize, we conduct extensive experiments on real-world datasets, results illustrate that incorporating price could improve the recommendation accuracy and our proposed PUP method outperforms other baselines on price-aware recommendation. We perform ablation study on price factor to confirm the critical role of price in e-commerce recommendation. Experiments on the quantization and fineness of the price factor illustrate the capability of our proposed method in capturing price preference. Our two-branch design disentangles the global and local effect of the price factor and serves great recommendation performance. Moreover, we investigate the consistency of price awareness across multiple categories. Furthermore, our proposed PUP shows superior performance when tackling cold-start problems.

%% file: tables/dataset.tex
\begin{table}[tb]
    \centering
    \caption{Statistics of the datasets}
    \label{tab::dataset}
    \begin{tabular}{|c|c|c|c|c|c|c|}
    \hline
    \textbf{Dataset} & \textbf{\#Users} & \textbf{\#Items} & \textbf{\#Cate} & \textbf{\#Price}  & \textbf{\#Interactions}
    \\
    \hline
    \hline
    Yelp & 20637 & 18907 & 89 & 4 & 505785
    \\
    \hline
    Beibei & 52767 & 39303 & 110 & 10 & 677065
    \\
    \hline
    \end{tabular}
    \vspace{-5pt}
\end{table}

%% file: tables/performance_comparison_ICDE_revision.tex
\begin{table*}[tb]
    \centering
    \caption{Top-K recommendation performance comparison on the Yelp and Beibei datasets (K is set to 50 and 100)}
    \label{tab::performance_comparison}
    \begin{tabular}{|c||c|c|c|c||c|c|c|c|}
    \hline
     & \multicolumn{4}{c||}{Yelp dataset} & \multicolumn{4}{c|}{Beibei dataset}
    \\
    \hline
    method & Recall@50 & NDCG@50 & Recall@100 & NDCG@100 & Recall@50 & NDCG@50 & Recall@100 & NDCG@100
    \\
    \hline
    \hline
    ItemPop & 0.0401 & 0.0182 & 0.0660 & 0.0247 & 0.0087 & 0.0027 & 0.0175 & 0.0046
    \\
    \hline
    BPR-MF & 0.1621 & 0.0767 & 0.2538 & 0.1000 & 0.0256 & 0.0103 & 0.0379 & 0.0129
    \\
    \hline
    \hline
    PaDQ & 0.1241 & 0.0572  & 0.2000 & 0.0767 & 0.0131 & 0.0056 & 0.0186 & 0.0068
    \\
    \hline
    FM & 0.1635 & \textbf{0.0771} & 0.2538 & 0.1001 & \textbf{0.0259} & 0.0104 & 0.0384 & 0.0130
    \\
    \hline
    \hline
    DeepFM &  0.1644 & 0.0769 & 0.2545 & 0.0998 & 0.0255 & 0.0090 & \textbf{0.0400} & 0.0122
    \\
    \hline
    GC-MC & 0.1670 & 0.0770 & \textbf{0.2621} & \textbf{0.1011} & 0.0231 & 0.0100 & 0.0343 & 0.0124
    \\
    \hline
    NGCF & \textbf{0.1679} & 0.0769 & 0.2619  & 0.1008 & 0.0256 & \textbf{0.0107} & 0.0383 & \textbf{0.0134}
    \\
    \hline
    PUP & \textbf{0.1765} & \textbf{0.0816} & \textbf{0.2715} & \textbf{0.1058} & \textbf{0.0266} & \textbf{0.0113} & \textbf{0.0403} & \textbf{0.0142}
    \\
    \hline
    impr.\% & 5.12\% & 5.84\% & 3.59\% & 4.65\% & 2.70\% & 5.61\% & 0.75\% & 5.97\%
    \\
    \hline
    \end{tabular}
\end{table*}

%% file: tables/ablation.tex
\begin{table}[tb]
    \centering
    \caption{Ablation study on the importance of price factor (topK=50)}
    \label{tab::ablation}
    \begin{tabular}{|c|c|c|c|c|}
    \hline
    method & Recall@50 & NDCG@50 & Recall@100 & NDCG@100
    \\
    \hline
    PUP w/o c,p & 0.0726 & 0.0211 & 0.1155 & 0.0285
    \\
    \hline
    PUP w/ c & 0.0633 & 0.0222 & 0.0944 & 0.0276
    \\
    \hline
    PUP w/ p & 0.0854 & 0.0277 & 0.1275 & 0.0350
    \\
    \hline
    PUP & 0.0890 & 0.0293 & 0.1336 & 0.0370
    \\
    \hline
    \end{tabular}
    \vspace{-5px}
\end{table}

%% file: tables/rank_uniform.tex
\begin{table}[tb]
    \centering
    \caption{Results of Different Quantization Process on Amazon Dataset}
    \label{tab::rank_uniform}
    \begin{tabular}{|c|c|c|c|c|}
    \hline
    Method & Recall@50 & NDCG@50 & Recall@100 & NDCG@100
    \\
    \hline
    Uniform & 0.0807 & 0.0264 & 0.1192 & 0.0331
    \\
    \hline
    Rank & 0.0885 & 0.0294 & 0.1313 & 0.0368
    \\
    \hline
    \end{tabular}
    \vspace{-10px}
\end{table}

%% file: tables/embedding_slicing.tex
\begin{table}[tb]
    \centering
    \caption{Performance comparison of different embedding size allocations on Yelp dataset (topk=50)}
    \label{tab::slicing}
    \begin{tabular}{|c|c|c|c|c|c|}
    \hline
    Allocation & 16/48 & 32/32 & 48/16 & 56/8 & 60/4
    \\
    \hline
    Recall & 0.1460 & 0.1689 & 0.1757 & \textbf{0.1765} & 0.1745
    \\
    \hline
    \end{tabular}
    \vspace{-5px}
\end{table}

%% file: tables/user_group_ICDE.tex
\begin{table}[tb]
    \centering
    \caption{performance comparison on different user groups on beibei dataset (metrics=NDCG@50)}
    \label{tab::entropy}
    \begin{tabular}{|c|c|c||c|}
    \hline
    user group & DeepFM & PUP & boost
    \\
    \hline
    consistent & 0.0091 & 0.0129 & 41.76\%
    \\
    \hline
    inconsistent & 0.0085 & 0.0086 & 1.18\%
    \\
    \hline
    \end{tabular}
    \vspace{-10px}
\end{table}

%% file: 6.relatedworks.tex
\section{Related Work}\label{Sec:RelatedWork}
\para{Price-aware Recommendation}
In e-commerce systems, price serves as a significant role in researches focusing on the task of revenue maximization. Nevertheless, price is seldom specifically utilized in improving recommendation. Schafer~\textit{et al.}~\cite{schafer1999recommender} first pointed out the potential of including price in e-commerce recommender systems. \davy{Wang~\textit{et al.}~\cite{wang2011utilizing} proposed $\mathrm{SVD_{\textit{util}}}$ which leveraged price to recommend items with higher net utility. }
Recently, several studies~\cite{KBS, WWW17, SIGIR14,umberto2015developing} started to approach price-aware recommendation, of which the aim is to better estimate users' preference with the help of products' price information. Asnat~\textit{et al.}~\cite{KBS} leveraged a Gaussian distribution to model users' \textit{acceptable price}, and when predicting user-item interactions, a user-price matching score calculated from the distribution is multiplied to the traditional user-item score to get the final predictions. A major concern is that directly matching user and price ignores the category-dependent influence of the price factor. Wang~\textit{et al.} ~\cite{WWW17} incorporate dynamic pricing of item and price elasticity (how the change of price affects user behaviors) into recommender systems via modeling users' decision making into three stages: category purchase, product choice and purchase quantity. However, in most real scenarios, decision making is always compressed to one stage because a user is directly exposed to item candidates selected by recommendation engines. Chen~\textit{et al.}~\cite{SIGIR14} applied collective matrix factorization (CMF)~\cite{CMF} and proposed a model called PaDQ to factorize user-item, user-pattern and item-price matrices simultaneously.
Umberto~\cite{umberto2015developing} explored the role of price in e-commerce recommendation and utilized price to improve business profit based on a multi-dimensional collaborative filtering approach~\cite{adomavicius2005incorporating}. In summary, these methods are not able to address the two main difficulties in integrating item price into recommender systems which have been discussed in previous sections.

\para{Attribute-aware Recommendation}
Attribute-aware recommendation is defined as to leverage user profiles or item descriptions to enhance recommendation~\cite{ContentRec}. Rendle~\textit{et al.}~\cite{FM} proposed Factorization Machines (FM) which estimates preferences with weighted sum of pairwise interactions between every two attributes (since 2-way FM is the most well known case). \davy{Juan~\textit{et al.}~\cite{juan2016field} further extended FM by adding field information which could distinguish among different feature interactions.} Cheng~\textit{et al.}~\cite{cheng2016wide} proposed a model named Wide\&Deep, which is the first work to utilize neural networks to capture various attributes in recommendation. He~\textit{et al.}~\cite{NFM} extended FM via replacing weighted sum with a Bi-Interaction layer and multi-layer perception. Guo~\textit{et al.}~\cite{DeepFM} extended FM with a deep neural network to model high order feature interactions in CTR prediction. Lian~\textit{et al.}~\cite{xDeepFM} combined compressed interaction network with multi-layer perceptions to learn both low- and high-order feature interactions. However, these models are relatively general models and not specifically designed for price-awareness modeling. Although these attribute-based recommendation models can be adapted to the task of price-aware recommendation by regarding price as an attribute, some intrinsic characteristics such as category-dependent price awareness could not be captured.